# Development and Evaluation of a Deep Neural Network for Histologic Classification of Renal Cell Carcinoma on Biopsy and Surgical Resection Slides

**The short running title: Deep Neural Network for classification of RCC**

Mengdan Zhu, MS[1], Bing Ren, MD[2], Ryland Richards, MD[2], Matthew Suriawinata[1], Naofumi Tomita, MS[1], Saeed Hassanpour, PhD[1,3,4*]

[1]Department of Biomedical Data Science, Geisel School of Medicine at Dartmouth, Hanover, NH 03755, USA

[2]Department of Pathology and Laboratory Medicine, Dartmouth-Hitchcock Medical Center, Lebanon, NH 03756, USA

[3]Department of Computer Science, Dartmouth College, Hanover, NH 03755, USA

[4]Department of Epidemiology, Geisel School of Medicine at Dartmouth, Hanover, NH 03755, USA

[*] Corresponding Author: Saeed Hassanpour, PhD
Postal address: One Medical Center Drive, HB 7261, Lebanon, NH 03756, USA
Telephone: (603) 650-1983
Email: Saeed.Hassanpour@dartmouth.edu

## Abstract

Renal cell carcinoma (RCC) is the most common renal cancer in adults. The histopathologic classification of RCC is essential for diagnosis, prognosis, and management of patients. Reorganization and classification of complex histologic patterns of RCC on biopsy and surgical resection slides under a microscope remains a heavily specialized, error-prone, and time-consuming task for pathologists. In this study, we developed a deep neural network model that can accurately classify digitized surgical resection slides and biopsy slides into five related classes: clear cell RCC, papillary RCC, chromophobe RCC, renal oncocytoma, and normal. In addition to the whole-slide classification pipeline, we visualized the identified indicative regions and features on slides for classification by reprocessing patch-level classification results to ensure the explainability of our diagnostic model. We evaluated our model on independent test sets of 78 surgical resection whole slides and 79 biopsy slides from our tertiary medical institution, and 69 randomly selected surgical resection slides from The Cancer Genome Atlas (TCGA) database. The average area under the curve (AUC) of our classifier on the internal resection slides, internal biopsy slides, and external TCGA slides is 0.98, 0.98 and 0.99, respectively. Our results suggest that the high generalizability of our approach across different data sources and specimen types. More importantly, our model has the potential to assist pathologists by (1) automatically pre-screening slides to reduce false-negative cases, (2) highlighting regions of importance on digitized slides to accelerate diagnosis, and (3) providing objective and accurate diagnosis as the second opinion.

## Introduction

Kidney cancer is among the ten most common cancers worldwide [1, 2], and approximately 90% of all kidney cancers are renal cell carcinoma (RCC) [3]. The classification of RCC consists of three major histologic RCC subtypes. Clear cell renal cell carcinoma (ccRCC) is the most common subtype (around 75% of all cases), papillary renal cell carcinoma (pRCC) accounts for about 15%-20% of RCC, and chromophobe renal cell carcinoma (chRCC) makes up approximately 5% of RCC [4]. The classic morphologic features of ccRCC include compact, alveolar, tubulocystic or rarely papillary architecture of cells with clear cytoplasm and characteristic network of small, thin-walled vessels [5]. Papillae or tubulopapillary architecture with fibrovascular cores and frequently with foamy macrophages are identified pRCC [6]. Of note, although renal oncocytoma is the most common benign renal tumor type, it still remains difficult to distinguish clinically from renal cell carcinoma including chRCC [7]. There are well-documented disparities in histologic appearances of chRCC and renal oncocytoma: chRCC shows prominent cell border, raisinoid nuclei and perinuclear halo, while oncocytoma displays nested architecture with myxoid or hyalinized stroma and cells with eosinophilic or granular cytoplasm and small round nuclei. However, the eosinophilic variant of chRCC can mimic the histologic features of oncocytoma, given their similar histogenesis [8].

   Histological classification of RCC is of great importance in patient's care, as RCC subtypes have significant implication in the prognosis and treatment of renal tumors [9-11]. Inspection and examination of complex RCC histologic patterns under the microscope, however, remain a time-consuming and demanding task for pathologists. The manual classification of RCC has shown a high rate of inter-observer and intra-observer variability [12], as renal tumors can have varied appearances and combined morphologic features, making classification difficult. With the advent of whole-slide images in digital pathology, automated histopathologic image analysis systems have shown great promise for diagnostic purposes [13-15]. Computerized image analysis has the advantage of providing a more efficient, objective, and consistent evaluation to assist pathologists in their medical decision-making processes. In recent years, significant progress has been made in applying deep learning techniques, especially convolutional neural networks (CNNs), to a wide range of computer vision tasks as well as biomedical imaging analysis applications [16-18]. CNN-based models can automatically process digitized histopathology images and learn to extract cellular patterns associated with the presence of tumors [19-21].

In this study, we developed a CNN-based model for classification of renal cell carcinoma based on surgical resection slides. We evaluated this model on 78 independent surgical resection slides from our tertiary medical institution and 69 surgical resection RCC slides from The Cancer Genome Atlas (TCGA) database. Furthermore, we evaluated this model for RCC classification on 79 biopsy slides from our institution. The study presented in this paper utilizes deep neural networks to automatically and accurately differentiate RCC from benign renal tumor cases and classify RCC subtypes on both surgical resection and biopsy slides.

## MATERIALS AND METHODS

### Data Collection

A total of 486 whole-slide images were collected from patients who underwent renal resection, including 30 normal slides with benign renal parenchyma and no renal neoplasm, from 2015 to 2019 from Dartmouth-Hitchcock Medical Center (DHMC), a tertiary medical institution in New Hampshire, USA. These hematoxylin and eosin (H&E) stained surgical resection slides were digitized by Aperio AT2 scanners (Leica Biosystems, Wetzlar, Germany) at 20× magnification (0.50 µm/pixel). We partitioned these slides into a training set of 385 slides, a development (dev) set of 23 slides, and a test set of 78 slides. Additionally, we collected 79 RCC biopsy slides from 2015 to 2017 from DHMC, as well as 69 whole-slide images of kidney cancer from TCGA for external validation. The use of the collected data was in accordance with the World Medical Association Declaration of Helsinki on Ethical Principles for Medical Research involving Human Subjects [22] and was approved by the Dartmouth-Hitchcock Human Research Protection Program (IRB). The distribution of whole-slide images that were used in this study is summarized in Table 1.

|  | Surgical Resection WSIs | | | | Biopsy WSIs |
|---|---|---|---|---|---|
|  | DHMC | | | TCGA | DHMC |
| Histologic Subtype | Training Set | Dev Set | Test Set #1 | Test Set #2 | Test #3 |
| normal | 15 | 5 | 10 | 9 | - |
| renal oncocytoma | 14 | 3 | 10 | - | 24 |
| chromophobe RCC | 15 | 5 | 18 | 20 | - |
| clear cell RCC | 285 | 5 | 20 | 20 | 34 |
| papillary RCC | 56 | 5 | 20 | 20 | 21 |
| Total | 385 | 23 | 78 | 69 | 79 |

**Table 1.** Distribution of the collected whole-slide images among renal cell carcinoma and benign subtypes. "-" indicates the corresponding subtype was not available in the dataset.

**Data Annotation**

Two pathologists (R.R. & B.R.) from the Department of Pathology and Laboratory Medicine at DHMC manually annotated the surgical resection whole-slide images in our training and development sets. In this annotation process, bounding boxes outlining regions of interest (ROIs) for each subtype were generated using Automated Slide Analysis Platform (ASAP), a fast viewer and annotation tool for high-resolution histopathology images [23]. Each ROI was associated and labeled as clear cell RCC, papillary RCC, chromophobe RCC, oncocytoma, or normal. All annotated ROIs were confirmed by one pathologist at a time before being broken into fixed-size patches for our model training and validation steps.

**Deep Neural Network for Patch Classification**

Given the large size of high-resolution histology images and the memory restrictions of currently available computer hardware, it is not feasible to analyze a whole-slide image all at once. Therefore, in this work, we use a computational framework[1] developed by our group that relies on deep neural network image analysis on small fixed-size patches with an overlap of 1/3 from the whole-slide images. These results are then aggregated through a confidence-based inference mechanism to classify the whole-slide images. As a result, this framework allows us to analyze a high-resolution, whole-slide image with a feasible memory requirement. Figure 1 shows the overview of our model in this study.

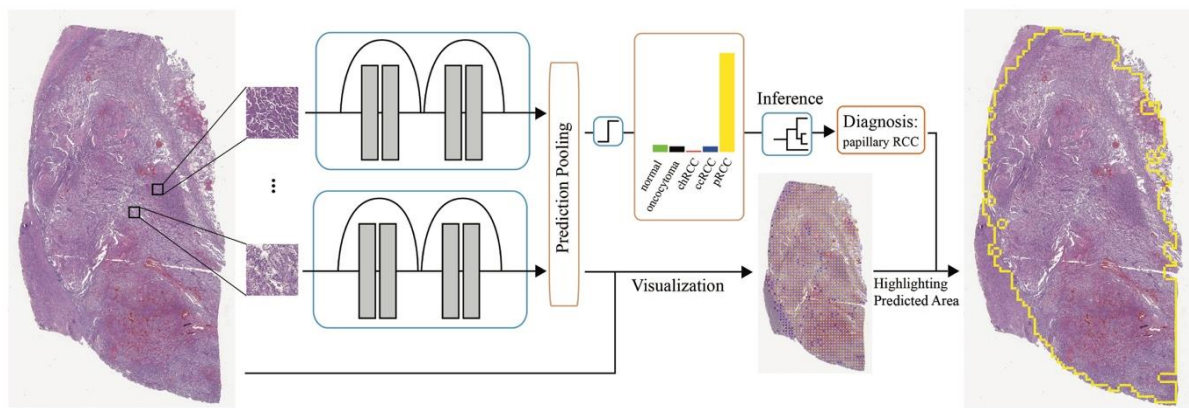

**Figure 1.** *Overview of our classification pipeline. Tissue patches are extracted from whole-slide images using a sliding-window method with 1/3 overlap after background removal. Deep neural networks extract histology features of the patches and compute patch-level confidence scores for each of the target classes. The patch-level predictions are filtered by low-confidence thresholding and aggregated by computing the percentage of patches that belong to each class in a whole-slide image. We classify a whole slide using a decision tree based on the computed percentages of each class. Patch predictions are also used for visualization, which illustrates the coverage of each class on slides.*

To do this, we utilized a sliding window approach [24] on the annotated ROIs in our training and development sets to generate fixed-size (i.e., 224×224 pixels) patches. To balance the dataset, we randomly selected the same number of 12,240 patches from the training set for each subtype. The distribution of this patch-level dataset is available in Table S1 in the Supplementary Material. We normalized the color intensity of patches and applied standard data augmentation methods, including random horizontal and vertical flips, random 90° rotations, and color jittering. For model training, we tried four variations of residual neural network (ResNet) architecture with different numbers of layers: ResNet-18, ResNet-34, ResNet-50 and ResNet-101. All the networks were initialized using He initialization [25]. These models used the multi-class cross entropy loss function and were trained for 40 epochs with an initial learning rate of 0.001. The learning rate was reduced by a factor of 0.9 every epoch during the training. The trained models assign a label with a confidence score (i.e., a prediction probability between 0 and 1) for each patch. We compared the trained models in our cross-validation process. Among the trained models, we selected a ResNet-18 model, which achieved the best average F1-score of 0.96 on the development set, for further whole-slide inference. The model's performance on the development set is summarized in Table S2 in the Supplementary Material.

---

[1] https://github.com/BMIRDS/deepslide

**Whole-slide Inference**

For whole-slide classification, our approach aggregated patch-level predictions based on their confidence scores. For each whole-slide image, we automatically processed the image by removing the white background, breaking down the remaining areas in each whole-slide image into fixed-size (i.e., 224×224 pixels) patches, and feeding the patches to our trained deep neural network to generate a pool of patch-level predictions. Of note, to enhance the robustness of our method, we removed all low-confidence patches from this pool so that their confidence scores were less than the threshold of 0.9. We performed a grid search to find the best threshold for the patch-level confidence score on the development set.

To aggregate the patch-level predictions, we computed the percentage of patches that belongs to each class in the pool of patches from a whole-slide image. We applied a grid-search optimization on patch-based statistics in the development set to build our inference criteria for whole-slide inference. In our whole-slide image inference criteria, if any of the renal subtypes (i.e., clear cell RCC, papillary RCC, chromophobe RCC, or oncocytoma) accounted for more than 5.0% of the total number of patches, we labeled the whole-slide image as a non-normal class with the greatest number of patches. Otherwise, we classified the whole-slide image as overall normal.

**Evaluation Metrics and Statistical Analysis**

To show the accuracy and generalizability of our approach, we evaluated our method on three different test sets: (1) 78 independent surgical resection whole-slide images from DHMC, (2) 69 surgical resection whole-slide images from the TCGA database, and (3) 79 biopsy whole-slide images from DHMC.

In this evaluation, we establish the gold standard for each whole-slide image in our test sets based on the original institutional label and the verification of a pathologist (R.R.) involved in our study. If there is any disagreement, we send the cases to our senior pathologist (B.R.) to resolve the disagreement. For this multi-class classification, we used precision, recall, the F1-score, and the area under the curve (AUC), as well as confusion matrices to show the discriminating performance of our approach for renal cancer classification. In addition, 95% confidence intervals (95% CIs) were computed using the bootstrapping method with 10,000 iterations for all the metrics [26].

**RESULTS**

Table 2 summarizes the per-class and average evaluation of our model on the first test set of surgical resection whole-slide images from DHMC. Our model achieved a mean accuracy of 0.97, a mean precision of 0.94, a mean recall of 0.92, a mean F1-score of 0.92, and a mean AUC of 0.98 on this internal test set of resection slides. Table 3 shows the performance summary of our model on randomly selected whole-slide images from the kidney renal carcinoma collection of the TCGA databases. We achieved high performance on these external resection whole-slide images with a mean accuracy of 0.98, a mean precision of 0.97, a mean recall of 0.95, a mean F1-score of 0.96, and a mean AUC of 0.99. Table 4 presents the per-class and mean performance metrics of our model on 79 biopsy whole-slide images from DHMC. Our model shows great generalizability on the internal biopsy test set, with a mean accuracy of 0.97, a mean precision of 0.97, a mean recall of 0.93, a mean F1-score of 0.95, and a mean AUC of 0.98.

| Subtype | Accuracy | Precision | Recall | F1-score | AUROC |
|---|---|---|---|---|---|
| normal | 1.00 (1.00-1.00) | 1.00 (1.00-1.00) | 1.00 (1.00-1.00) | 1.00 (1.00-1.00) | 1.00 |
| oncocytoma | 0.97 (0.95-1.00) | 1.00 (1.00-1.00) | 0.80 (0.63-0.95) | 0.89 (0.77-0.97) | 0.97 |
| chromophobe RCC | 0.94 (0.90-0.97) | 0.93 (0.84-1.00) | 0.78 (0.65-0.89) | 0.85 (0.76-0.92) | 0.98 |
| clear cell RCC | 0.97 (0.95-1.00) | 0.91 (0.83-0.98) | 1.00 (1.00-1.00) | 0.95 (0.91-0.99) | 0.98 |
| papillary RCC | 0.96 (0.94-0.99) | 0.87 (0.78-0.95) | 1.00 (1.00-1.00) | 0.93 (0.88-0.97) | 0.99 |
| Average | 0.97 (0.95-0.98) | 0.94 (0.91-0.97) | 0.92 (0.87-0.95) | 0.92 (0.88-0.96) | 0.98 |

**Table 2.** Model's performance on 78 surgical resection whole-slide images in our independent test set from DHMC. The 95% confidence interval is also included for each measure.

| Subtype | Accuracy | Precision | Recall | F1-score | AUROC |
|---|---|---|---|---|---|
| normal | 1.00 (1.00-1.00) | 1.00 (1.00-1.00) | 1.00 (1.00-1.00) | 1.00 (1.00-1.00) | 1.00 |
| chromophobe RCC | 1.00 (1.00-1.00) | 1.00 (1.00-1.00) | 1.00 (1.00-1.00) | 1.00 (1.00-1.00) | 1.00 |
| clear cell RCC | 0.96 (0.93-0.98) | 0.95 (0.88-1.00) | 0.90 (0.82-0.97) | 0.92 (0.87-0.97) | 0.96 |
| papillary RCC | 0.96 (0.93-0.98) | 0.95 (0.88-1.00) | 0.90 (0.82-0.97) | 0.92 (0.87-0.97) | 0.98 |
| Average | 0.98 (0.96-0.99) | 0.97 (0.94-1.00) | 0.95 (0.91-0.98) | 0.96 (0.93-0.98) | 0.99 |

**Table 3.** Model's performance metrics and their 95% confidence intervals on 69 randomly selected surgical resection whole-slide images from the TCGA database.

| Subtype | Accuracy | Precision | Recall | F1-score | AUROC |
|---|---|---|---|---|---|
| oncocytoma | 0.96 (0.94-0.99) | 1.00 (1.00-1.00) | 0.87 (0.79-0.95) | 0.93 (0.88-0.98) | 1.00 |
| clear cell RCC | 0.96 (0.94-0.99) | 1.00 (1.00-1.00) | 0.91 (0.85-0.97) | 0.95 (0.92-0.98) | 0.95 |
| papillary RCC | 0.97 (0.95-1.00) | 0.91 (0.83-0.98) | 1.00 (1.00-1.00) | 0.95 (0.91-0.99) | 0.99 |
| Average | 0.97 (0.94-0.99) | 0.97 (0.94-0.99) | 0.93 (0.88-0.97) | 0.95 (0.90-0.98) | 0.98 |

**Table 4.** Model's performance metrics and their 95% confidence intervals on 79 biopsy whole-slide images from DHMC.

The confusion matrices for each of our three test sets are shown in Figure 2. Overall, the normal cases could be easily recognized by our model, whereas a minor portion of oncocytoma cases could be misclassified as chromophobe RCC and papillary RCC in both surgical resection cases and biopsy cases. We provide a detailed error analysis in the discussion section. The Receiver Operating Characteristic (ROC) curves of all the test sets are plotted in Figure 3.

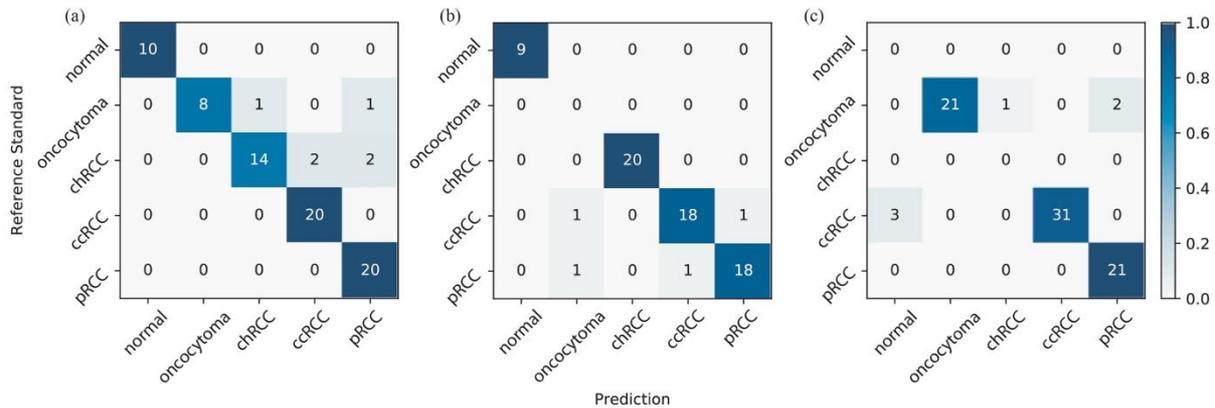

**Figure 2.** *Each confusion matrix compares the classification agreement of our model with pathologists' consensus for each of our three test sets: (a) surgical resection whole-slide images from DHMC, (b) surgical resection whole-slide images from TCGA, and (c) biopsy whole-slide images from DHMC.*

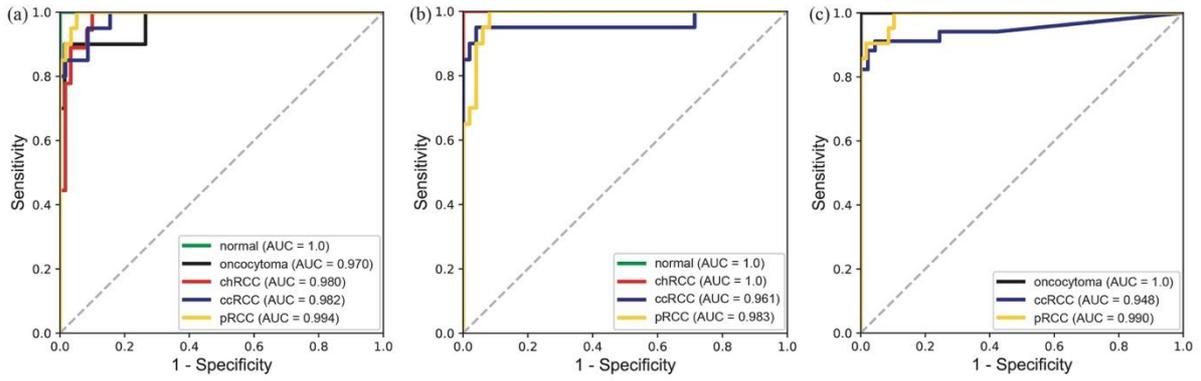

**Figure 3.** *Receiver operating characteristic curves for (a) surgical resection whole-slide images from DHMC, (b) surgical resection whole-slide images from TCGA, and (c) biopsy whole-slide images from DHMC.*

We visualize the patches on whole-slide images in our test sets with a color-coded scheme according to the classes predicted by our model. This visualization provides insights into the major regions and features that contribute to the classification decisions of our method, to avoid the "black-box" approach toward the outputs. Figure 4 shows a sample visualization for slides from each test set. More visualization examples from the DHMC surgical resection test set are included in Figure S1 in the Supplementary Material.

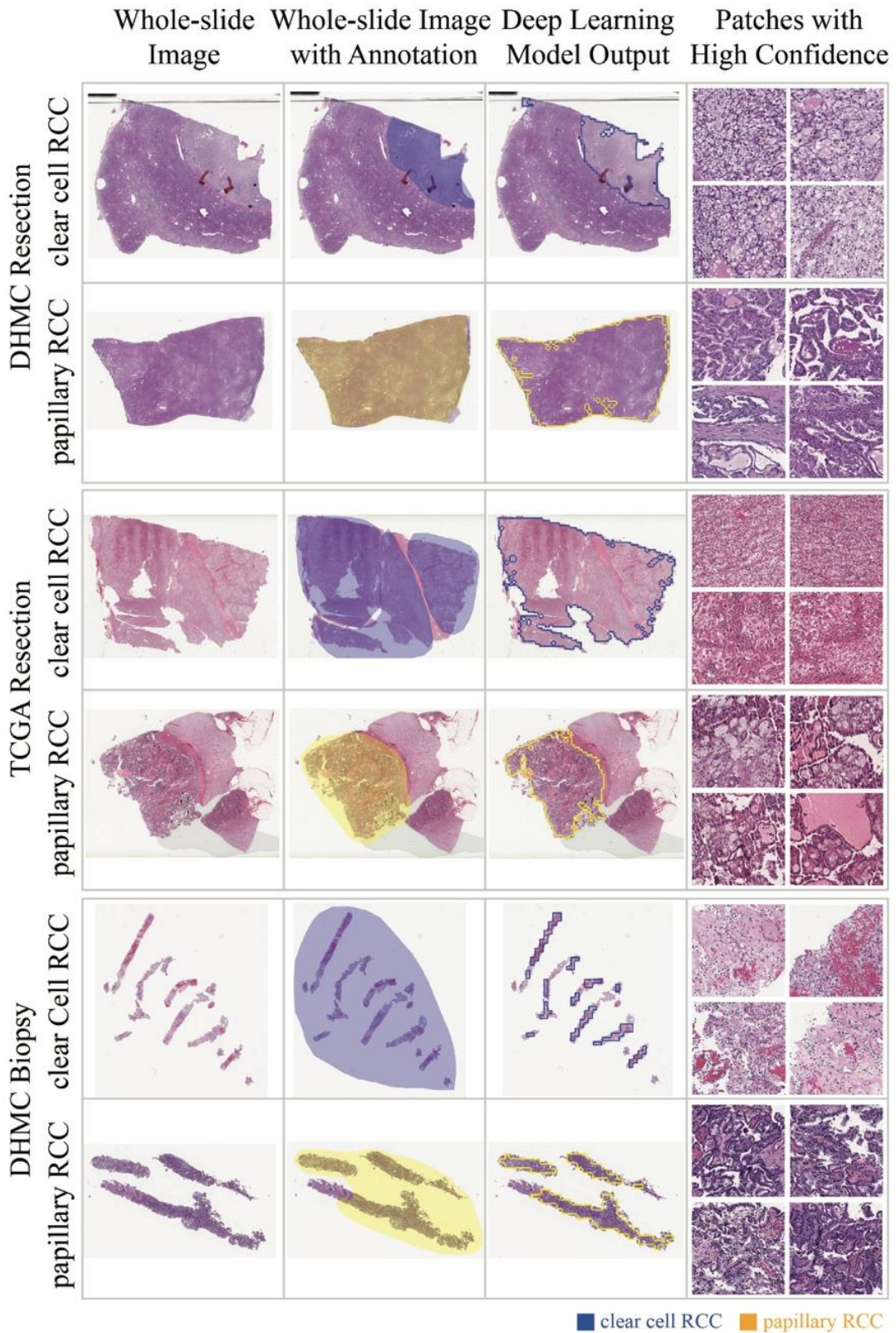

**Figure 4.** *Examples of visualized slides from our test sets with highlighted regions of interest for predicted classes using our model. Clear cell RCC and papillary RCC classes are common among the three test sets and thus are used for this illustration. Top row: A surgical resection whole-slide image in the DHMC test set. Middle row: A surgical resection whole-slide image from the TCGA test set. Bottom row: A biopsy whole-slide image from DHMC.*

**DISCUSSION**

Classification of renal cell carcinoma subtype is a clinically important task that enables clinicians to predict prognosis and to choose the optimal management for patients with RCC. Different RCC subtypes may have different prognosis, underlining the importance of differentiation of these subtypes. Clear cell RCC has a worse prognosis compared to chromophobe or papillary RCC at the same stage [27-30]. Of note, the most common benign renal tumor is oncocytoma (3-7% of all renal tumors) and is known for mimicking RCC on histology slides [31]. Therefore, it is very important to recognize different subtypes of RCC as well as benign renal neoplasms such as oncocytoma.

This study proposed and evaluated a deep neural network model for automated renal cell carcinoma classification on both surgical resection and biopsy whole-slide images. We chose ResNet-18 architecture as the backbone of our pipeline, which involved a patch-prediction aggregation strategy. Our final model achieved an average F1-score of 0.92, 0.96, and 0.95 for independent resection whole-slide image test sets from DHMC and TCGA databases, and DHMC biopsy whole-slide images, respectively. This study is the first step toward utilizing deep learning methods to automatically classify RCC subtypes and oncocytoma on histopathology images.

Notably, our model achieved higher performance on the test set of TCGA resection slides in comparison to our independent internal test set from DHMC. We have conducted the Mann Whitney U test on the model's performance on the test set of DHMC resection slides and the test set of TCGA resection slides. The p-value of this comparison is not significant at the level of 0.05 statistical significance. The difference in performance on those test sets could be due to the variations among the compilation processes of the datasets. All test sets were carefully and independently reviewed by two collaborator pathologists based on their qualitative assessment our internal resection test set, which included challenging cases that would require high subspecialty expertise, while the cases of the TCGA dataset, which were collected from various institutions, were more straightforward to classify. We suspect the TCGA collection process included an inherent pre-filtering mechanism to remove some edge cases and share high-quality classic cases publicly. In addition, oncocytoma was not included in TCGA data set. Therefore, the TCGA dataset may contain more typical histopathologic patterns, which are easier for our model to identify. That said, the higher performance metrics on the external public dataset in our evaluation illustrate the strong generalizability of our approach for differentiating and classifying RCC subtypes.

Previous work on machine learning applications to kidney cancer histopathology is mostly focused on resection slides and three RCC subtypes, without the consideration of benign or oncocytoma classes, with validation on a single test set [17, 32-34]. Recently, a combination of a convolutional neural network and Directed Acyclic Graph -support vector machine (DAG-SVM) was used for classification of three RCC subtypes using the TCGA dataset [17]. Our study stands out from this previous study for several reasons: (1) our approach follows a more intuitive methodology based on patch-level confidence scores and achieved a higher performance on the dataset; (2) our method was evaluated on both DHMC and TCGA datasets to show its generalizability on surgical resection whole-slide images; (3) our study includes identification of benign renal neoplasm, oncocytoma, in addition to all major RCC subtypes; and (4) we showed the application and generalizability of our model to a test set of biopsy whole-slide images, which also achieved promising results.

Of note, because the TCGA dataset is focused on malignant cancer cases, oncocytoma, a benign subtype, does not exist among the TCGA whole-slide images. Therefore, this subtype was not included in the surgical resection slides in our external test set. Additionally, chromophobe RCC makes up about 5% of RCC occurrences and we could only identify a few chromophobe biopsy slides at DHMC. Similarly, for clinical purposes, only a few normal biopsy slides are stored at DHMC, as more emphasis is put on renal tumor biopsy slides. Considering the prevalence and availability of chromophobe biopsy slides, and the availability of normal slides at our institution, we excluded chromophobe RCC and normal class from our biopsy test set, and evaluated our model on the two major RCC subtypes (i.e., clear cell RCC and papillary RCC) and the major renal benign tumor type (i.e., renal oncocytoma). Notably, the generalizability of our model to biopsy whole-slide images has a wide range of application, as it could assist clinicians with fast and reliable diagnoses and follow-up recommendations for patients.

Manual histopathological analysis is a tedious and time-consuming task that could induce errors and variability among different pathologists. Our model addresses this limitation by providing a new technology that has the potential to help pathologists achieve a more efficient, objective, and accurate diagnosis and classification of renal cell carcinoma. In particular, our approach could provide clinical assistance and a second opinion to general surgical pathologists that are not specialized in genitourinary pathology.

Our error analysis shown in Figure S2 in the Supplementary Material demonstrates that the misclassifications of our model are mainly due to the atypical morphologic patterns in the histopathologic images. In the DHMC resection test set, chromophobe RCC is misclassified as clear cell RCC because of the substantial clear cytoplasm and thin walled vasculature in the images (Figure S2a). Oncocytoma is misclassified as chromophobe RCC or papillary RCC due to focal tubular growth pattern and less characteristic stroma present (Figure S2b). In the TCGA test set, papillary RCC is misclassified as clear cell RCC due to focal tumor cells with clear cytoplasm and thin walled vasculature (Figure S2c) and clear cell RCC is misclassified as papillary RCC due to focal papillary formation and less clear cytoplasm (Figure S2d).

As a future direction, we plan to expand our dataset and test sets through external collaborations for a more robust and extensive evaluation of RCC subtypes. According to our error analysis, one of our model's limitations is the misclassification of clear cell RCC as normal in the biopsy slides. To address this limitation, we will pursue developing an adaptive thresholding method that is attentive to differences between biopsy slides and resection slides. Moreover, recent studies suggest that weakly supervised learning frameworks with attention mechanisms are effective for whole slide classification, which we can utilize in conjunction with our fully-supervised learning method to further improve the accuracy of our model [35, 36]. Finally, we plan to implement a prospective clinical trial to validate this approach in clinical settings and quantify its impact on the efficiency and accuracy of pathologists' diagnosis of renal cancer.

## ACKNOWLEDGMENTS

The authors would like to thank Lamar Moss for his help and suggestions to improve the manuscript.

## COMPETING INTERESTS

None Declared.

## FUNDING

This research was supported in part by grants from the US National Library of Medicine (R01LM012837) and the US National Cancer Institute (R01CA249758).

**Supplementary Material**

**Figure S1: Typical Examples of Visualized Slides using the Deep Learning Model**

**Figure S2: Error Analysis**

**Table S1: The Distribution of the Patch-level Development Dataset**

**Table S2: Model Performance in the Patch-level Development Set**

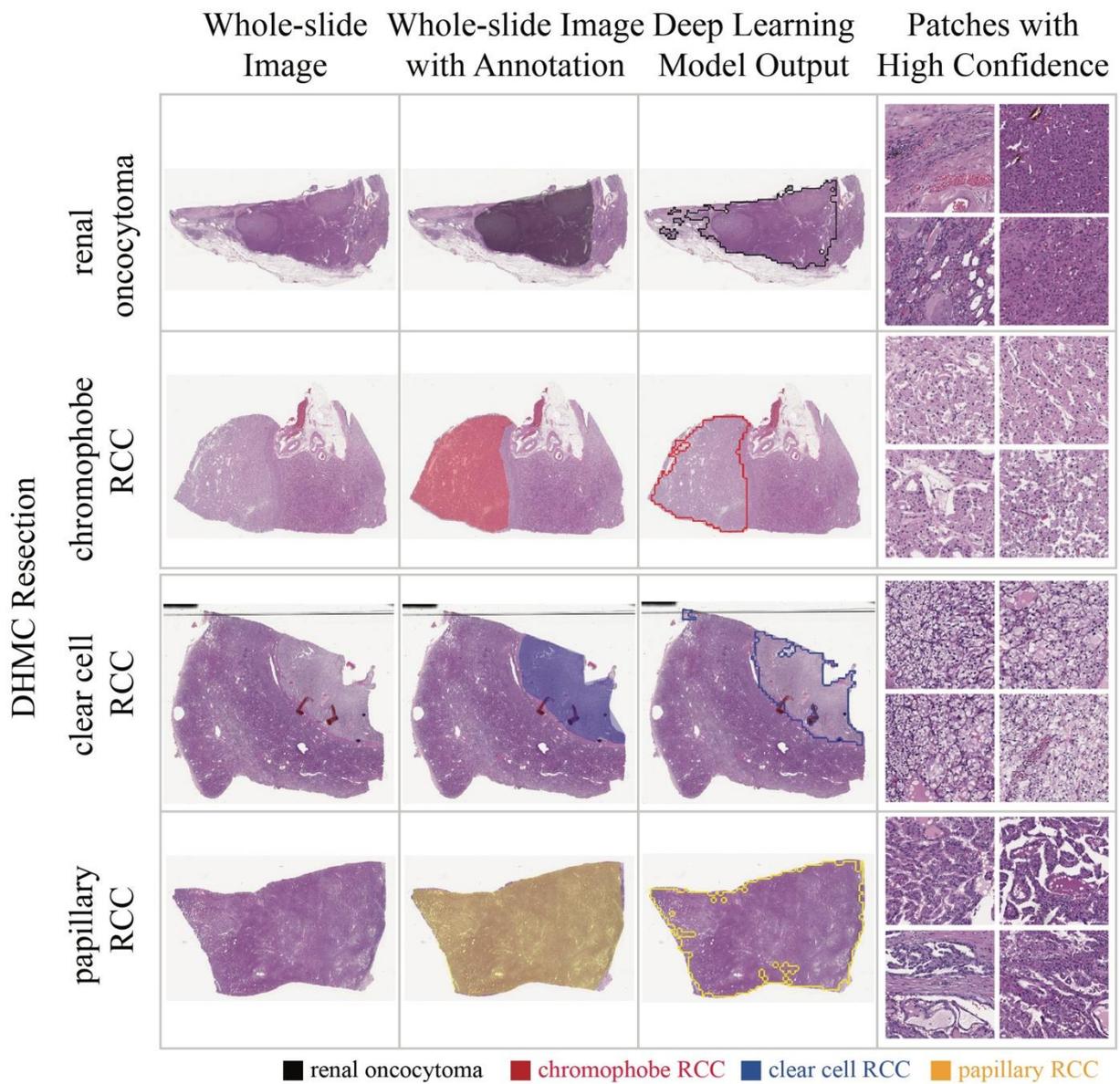

**Figure S1: Typical Examples of Visualized Slides using the Deep Learning Model.**
*Examples of visualized slides in our DHMC test set with highlighted regions for predicted classes. Each example slide for a subtype is presented with pathologists' annotations of non-normal areas, model output, and patches with the model's high confidence.*

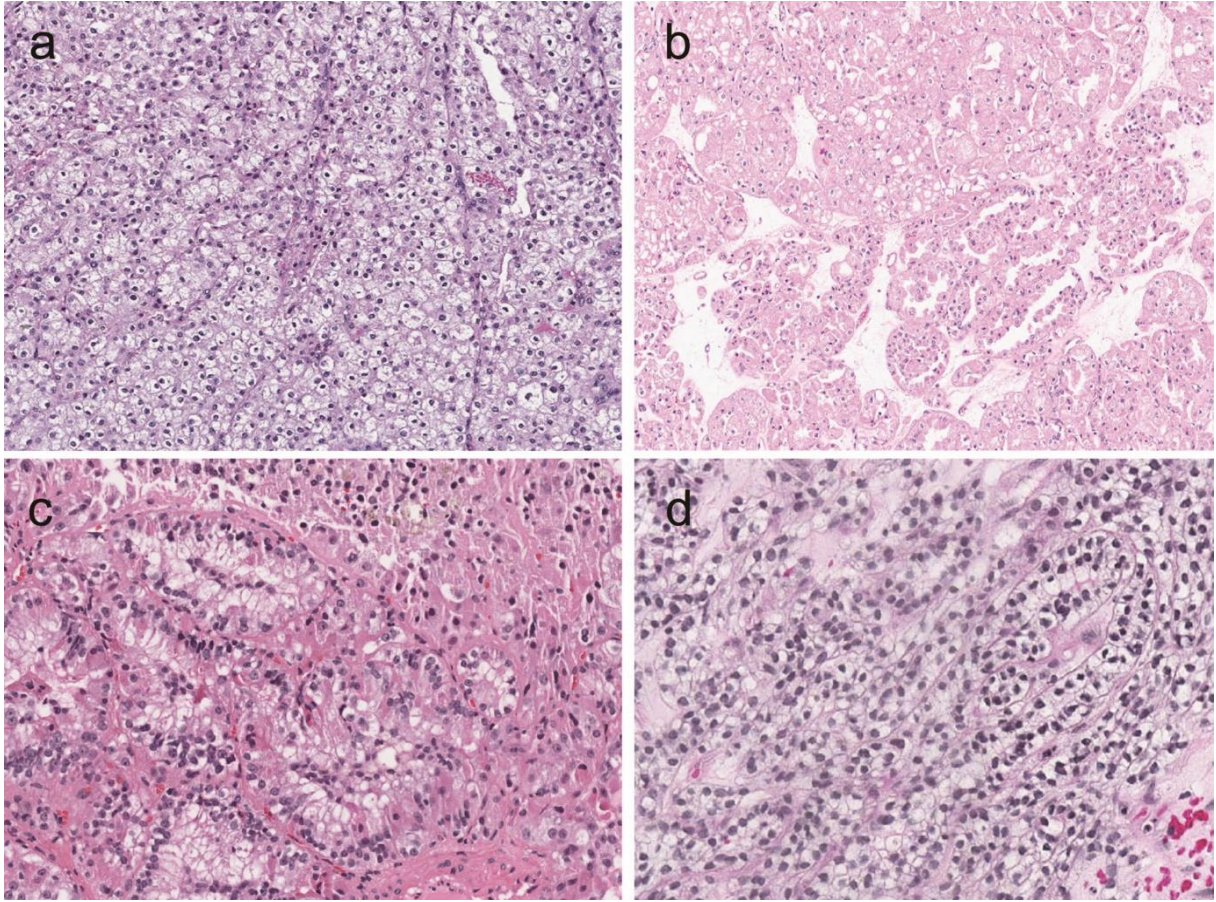

**Figure S2: Error Analysis.** *(a) The misclassification of chromophobe RCC as clear cell RCC or papillary RCC because of the substantial clear cytoplasm or papillary structure with fibrovascular cores. (b) The misclassification of oncocytoma as chromophobe RCC due to focal tubular growth pattern and less characteristic stroma present. (c) The misclassification of papillary RCC as clear cell RCC due to focal tumor cell with clear cytoplasm and thin walled vasculature. (d) The misclassification of clear cell RCC as papillary RCC due to the mimic of focal papillary formation and less clear cytoplasm.*

**Table S1: The Distribution of the Patch-level Development Dataset**

| Subtype | WSI | Patches |
| --- | --- | --- |
| normal | 5 | 7,639 |
| renal oncocytoma | 3 | 5,478 |
| chromophobe RCC | 5 | 12,308 |
| clear cell RCC | 5 | 10,989 |
| papillary RCC | 5 | 8,385 |
| Total | 23 | 44,799 |

**Table S2: Model performance on the patch-level Development set**

| Subtype | Precision | Recall | F1-score |
| --- | --- | --- | --- |
| normal | 1.00 | 1.00 | 1.00 |
| renal oncocytoma | 1.00 | 1.00 | 1.00 |
| chromophobe RCC | 1.00 | 1.00 | 1.00 |
| clear cell RCC | 0.83 | 1.00 | 0.91 |
| papillary RCC | 1.00 | 0.80 | 0.89 |
| Average | 0.97 | 0.96 | 0.96 |